\documentclass[12pt,preprint]{aastex}

\def\msun{{\rm M_\odot}}
\def\msuny{{\rm M_\odot yr^{-1}}}

\shorttitle{On the Classification of Classical Novae}
\shortauthors{Jos\'e et al.}

\begin{document}

\title{The Impact of the Chemical Stratification of White Dwarfs on the
       Classification of Classical Novae}

\author{Jordi Jos\'e}
\affil{Departament de F\'\i sica i Enginyeria Nuclear,
      Universitat Polit\`ecnica de Catalunya, Av. V\'{\i}ctor Balaguer, s/n, E-08800 Vilanova i la Geltr\'u
      (Barcelona),  and Institut d'Estudis Espacials de Catalunya (IEEC-UPC), 
      Ed. Nexus-201, C/ Gran Capit\`a 2-4, E-08034, Barcelona, Spain.}
\email{jordi.jose@upc.es}

\author{Margarita Hernanz}
\affil{Institut de Ci\`encies de l'Espai (CSIC), and Institut 
        d'Estudis Espacials de Catalunya (IEEC-CSIC), 
      Ed. Nexus-201, C/ Gran Capit\`a 2-4, E-08034, Barcelona, Spain.}
\email{hernanz@ieec.fcr.es}

\author{Enrique Garc\'\i a--Berro}
\affil{Departament de F\'\i sica Aplicada,
      Universitat Polit\`ecnica de Catalunya, 
      Av. del Canal Ol\'\i mpic, s/n, E-08860, Castelldefels (Barcelona)
      and Institut d'Estudis Espacials de Catalunya (IEEC-UPC), 
      Ed. Nexus-201, C/ Gran Capit\`a 2-4, E-08034, Barcelona, Spain.}
\email{garcia@fa.upc.es}

\and

\author{Pilar Gil--Pons}
\affil{Departament de F\'\i sica Aplicada,
      Universitat Polit\`ecnica de Catalunya, 
      Av. del Canal Ol\'\i mpic, s/n, E-08860, Castelldefels (Barcelona), Spain.}
\email{pilar@fa.upc.es}

\begin{abstract}
 We analyse the impact of the initial abundances 
  of the underlying white dwarf core on the nucleosynthesis
 accompanying classical nova outbursts, in the framework of hydrodynamic
 models of the explosion. Specifically, we take into account the
 chemical stratification of the white dwarf. It turns out that
 the presence of a thick CO-buffer on top of the ONe-rich core, as results
 from detailed calculations of previous evolution of the
 progenitor star, may lead to significant amounts of both $^7$Li and $^{26}$Al,  
  after an outburst that, due to the lack of neon isotopes in the ejecta, 
 would be misclassified as a non-neon nova (i.e., CO nova).
\end{abstract}

\keywords{novae, cataclysmic variables -- nucleosynthesis, abundances -- white dwarfs}

\section{Introduction}

 Classical nova outbursts are powered by thermonuclear runaways (hereafter, TNRs) that
 take place in the accreted envelopes of white dwarfs in close binary systems
 (Starrfield et al. 1972). Observationally, classical novae are classified into different 
 categories, based on detailed analysis of the 
 decline of the light curve (fast versus slow novae), and also on the chemical pattern
 spectroscopically inferred from the ejecta (CO versus ONe novae). 
 Atomic abundance determinations from the ejecta of several well-observed novae have
 provided a very useful tool to constrain theoretical models of the explosion.  
 Significant progress was achieved after the discovery of the large metallicities that characterize
 the nova ejecta, with values ranging from 0.043 for Nova RR Pic 1925 (Williams \& Gallagher
 1979) to 0.86 for Nova V1350 Aql 1982 (Andr\"ea et al. 1994; Snijders et al. 1987).
 In order to account for this effect, the modeling of the explosion required 
 some mixing between the solar-like material (transferred from the large, Main
 Sequence companion on top of the white dwarf by Roche lobe overflow) and the
 outermost layers of the underlying white dwarf. The reason is twofold:
 on one hand, at the typical temperatures expected during the course of the TNR, the amount 
 of leakage from the CNO cycle is very limited (Jos\'e \& Hernanz 1998), and hence, the 
 observed abundances of elements ranging from Ne to Ca (significantly overproduced with
 respect to solar proportions in some novae) cannot be explained in a natural way in terms
 of nuclear burning processes. Secondly, models computed with an envelope of solar
 metallicity are unable to reproduce the {\it gross} observational features of classical novae.
 Several mechanisms have been proposed to account for this mixing, including
 diffusion induced convection (Prialnik \& Kovetz 1984; Kovetz \& Prialnik 1985; Iben,
 Fujimoto \& MacDonald 1991, 1992; Fujimoto \& Iben 1992), shear mixing (MacDonald 1983, Livio
 \& Truran 1987),  convective overshoot induced flame propagation (Woosley 1986) or
 convection induced shear mixing (Kutter \& Sparks 1989).  Several multidimensional
 effects have also been tested with the recent development of two- and/or three-dimensional
 codes (Glasner, Livne \& Truran 1997; Kercek, Hillebrandt \& Truran 1998, 1999; Glasner \& Livne 2002),
 including  mixing by non-gravity wave breaking on white dwarf surfaces (Rosner et al. 2001; 
 Alexakis et al. 2003).

 Despite of the efforts carried out so far in the search for this mixing mechanism, 
 there is not yet a satisfactory and self-consistent explanation capable to account for the wide
 spread in metallicities inferred from nova ejecta. So far, the usual assumption is to arbitrarily parametrize the 
 mixing at the onset of the accretion phase, guided by observational
 grounds (Politano et al. 1995).  Either through this crude approach or by means of a real physical mechanism,
 yet unknown, one of the important ingredients in the mixing involves the nature of
 the underlying white dwarf star. 
 Moreover, the pioneeing discovery of Ne in the ejecta of Nova V1500 Cygni 1975 (Ferland \& Shields 1978; 
 Lance et al. 1988), Nova V1370 Aql 1982 (Snijders et al. 1984, 1987; Andr\"ea et al. 1994), 
 Nova V693 CrA 1981 (Williams et al. 1985; Andr\"ea et al. 1994; Vanlandingham et al. 1997)
 and Nova QU Vul 1984 (Andrillat \& Houziaux 1985; Gehrz, Grasdalen \& Hackwell 1985; 
 Andr\"ea et al. 1994; Saizar et al. 1992; Austin et al. 1996) was of significant importance,
 since it clearly pointed out the existence of two classes of nova outbursts: those 
 showing large amounts of Ne in the spectra, overabundant with respect to the solar
 value, and those mainly enriched in C and O 
 (see Table 2 in Gehrz et al. (1998) for a
 large sample of mass fractions inferred from optical and ultraviolet spectroscopy). 
 This dicotomy essentially results from stellar 
 evolution: whereas CO white dwarfs are
 the remnants of the evolution of intermediate-mass stars (M $\leq 8-10 \msun$) that undergo 
 hydrogen and helium burning (leaving a compact object that is C- and O-rich), slightly more 
 massive stars (up to $\sim 11 \msun$) undergo another evolutionary phase, carbon burning, leaving
 an ONe-rich object instead.  The reader is referred to Starrfield, Sparks \& Truran (1986) for the 
 first hydrodynamic models of novae involving ONe(Mg) white dwarfs. 

In the last decades, many efforts have been devoted to outline the nucleosynthesis accompanying nova
outbursts (see Jos\'e 2002, and references therein). In fact, a direct
 comparison between models and observations tells us that the overall nucleosynthetic picture is reasonably
 well understood: not only the endpoint for nova nucleosynthesis (around Ca) but
 also the atomic abundances inferred from the ejecta (Gehrz et al. 1998) are well reproduced (Jos\'e \&
 Hernanz 1998), with no major flaws, despite of the long-standing problems
  associated with the modeling of the explosion. 
 
 In this Letter, we analyze the impact of the specific choice of the chemical pattern of the outermost layers
 of massive (ONe) white dwarf cores in the composition of the ejecta, a subject that has deserved no particular attention 
 in the literature despite of its importance. 

\begin{deluxetable}{lrrr}
\tablecaption{Results of the evolution of 1.25 $\msun$ ONe white dwarfs. \label{tbl-1}}
\tablewidth{0pt}
\tablehead{
\colhead{    } & \colhead{A}   & \colhead{B}   & \colhead{C} 
}
\startdata
X($^{12}$C)$\rm _{initial}$      & 6.1 (-3) &  6.0 (-2) &  0.23     \\
X($^{16}$O)$\rm _{initial}$      & 0.26     &  0.28     &  0.26     \\
X($^{20}$Ne)$\rm _{initial}$     & 0.16     &  0.10     &  8.1 (-4) \\
\hline
$\rm \Delta M_{env}$ ($10^{-5} \msun$) & 2.20&  1.54     &   1.23    \\
$\rm t_{rise}$ ($10^5$ s)      &  121     &  33.7     &   2.54    \\
$\rm t_{max}$ (s)              &  313     &  134      &    45     \\
$\rm T_{peak}$ ($10^8$ K)        &  2.51    &  2.31     &   2.22    \\
$\rm K_{ejec}$ ($10^{45}$ ergs s$^{-1}$)& 1.52& 1.13  &   1.04    \\
$\rm \Delta M_{ejec}$ ($10^{-5} \msun$) & 1.79& 1.25     &  1.00     \\
X($^1$H)                     & 0.28     &  0.29     &  0.29     \\
X($^4$He)                    & 0.22     &  0.21     &  0.18     \\
X($^7$Li)                    & 7.7 (-7) &  8.4 (-6) &  1.1 (-5) \\
X($^{12}$C)                  & 2.4 (-2) &  3.0 (-2) &  4.7 (-2) \\
X($^{13}$C)                  & 3.3 (-2) &  4.1 (-2) &  7.3 (-2) \\ 
X($^{14}$N)                  & 3.9 (-2) &  3.9 (-2) &  1.1 (-1) \\
X($^{15}$N)                  & 4.3 (-2) &  5.2 (-2) &  6.8 (-2) \\
X($^{16}$O)                  & 6.8 (-2) &  1.3 (-1) &  1.9 (-1) \\
X($^{17}$O)                  & 5.5 (-5) &  5.5 (-5) &  5.5 (-5) \\
X($^{18}$F)\tablenotemark{a} & 2.8 (-4) &  3.4 (-4) &  3.0 (-4) \\
X($^{20}$Ne)                 & 1.7 (-1) &  1.2 (-1) &  9.4 (-4) \\
X($^{22}$Na)                 & 3.3 (-4) &  2.2 (-4) &  9.3 (-7) \\
X($^{26}$Al)                 & 6.0 (-4) &  6.0 (-4) &  3.7 (-4) \\
X($^{28}$Si)                 & 5.8 (-2) &  4.4 (-2) &  6.6 (-3) \\
X($^{29}$Si)                 & 1.1 (-3) &  4.8 (-4) &  4.4 (-5) \\
X($^{30}$Si)                 & 6.3 (-3) &  1.3 (-3) &  4.7 (-5) \\
\enddata

\tablenotetext{a}{The mass fraction of the radioactive isotope $^{18}$F is given 
                  at 40 minutes after $\rm T_{peak}$.}

\end{deluxetable}

\section{Results}

 Contrary to the usual assumption, that adopts as the composition
 of the outermost layers of the white dwarf that of the 'bare' ONe core, we take here into account  
 the chemical stratification of the interior. More specifically, we estimate
 the role played by the presence of a thick CO-buffer that remains on top of the ONe-rich core,
 according to the recent stellar evolutionary models computed by Garc\'\i a--Berro, Ritossa, \& Iben (1997), for
 stars in the mass range $8 \leq$ M/$\msun$ $< 11$. Such stars form an ONe core, after burning C in a
 partially degenerate core.
 The chemical structure of the abovementioned CO-buffer is shown in Fig. 1 (right panel):  
 it is characterized by a very flat profile with similar amounts of $^{12}$C and $^{16}$O, plus
 traces of $^{25}$Mg and $^{22}$Ne. It is worth noticing that although all models within this mass
 range end up with a CO-buffer on top of the ONe core, its size depends strongly on the specific
 value of the mass of the progenitor at the Main Sequence (see, for instance, Gil--Pons \& 
 Garc\'\i a--Berro 2001, 2002; Gil--Pons et al. 2003).
 Nevertheless, for the scope of this paper, it is enough to say that the C/O fraction in such CO-buffers 
 is rather independent of the mass of the progenitor. Hence, the structure displayed in Fig. 1,
 corresponding to the evolution of a single 9 $\msun$ star (Garc\'\i a--Berro et al. 1997) can be regarded
 as representative. 
Another common issue for all models in this mass range is the existence of 
 a transition zone that connects the CO-buffer and the ONe-rich core of the white dwarf,
 through a region (Fig. 1, middle panel) characterized by steep chemical profiles 
(see, for instance, $^{12}$C, $^{20}$Ne, or $^{24}$Mg). The pattern shown in Fig. 1 (middle panel), 
corresponds indeed to a typical abundance profile of the trasition zone, for stars in 
this mass range.
 
 It is not clear whether this CO-rich buffer (and/or the transition zone) is lost in the late stages of 
 the evolution of the star or rather, if accretion settles on top of this layer, 
 a possibility that has been ignored in previous works. 
 In this Letter, we analyse this issue by performing a series of hydrodynamic calculations of classical nova 
 outbursts, and discuss its impact 
  on the resulting explosion as well as on the accompanying nucleosynthesis. 
 Calculations have been performed with the 
 {\it Shiva} code, a spherically-symmetric, implicit, hydrodynamic code, in Lagrangian
formulation (see Jos\'e \& Hernanz 1998, for details) that follows the course of the explosion
from the onset of accretion up to the explosion and ejection stages. The calculations 
presented here are similar to those published in previous papers 
 except for the chemical composition adopted for the envelope. Here, we assume that 
the solar-like accreted material is mixed either with material from the CO-buffer (hereafter, 
Model C), or with material from the transition zone (Model B). It is worth
noticing that the second case implicitly assumes that the former CO-buffer has been lost during previous evolution.
Both models assume that the TNR takes place on top of a $1.25 \msun$ white dwarf that is accreting
material from a Main Sequence companion at a rate of $2\times 10^{-10} \msuny$. 
 As in previous papers (see discussion in Jos\'e
 \& Hernanz 1998), and due to the lack of a fully self-consistent mechanism of mixing, capable to account for
metallicity enrichments as high as 0.86 by mass, we adopt the usual approach of an arbitrary mixing ratio
between the solar-like accreted material and the outermost layers of the white dwarf at the onset of accretion
(see Politano et al. 1995).
For the calculations reported in this paper a 50\% mixing ratio (representative of the average metallicity
inferred from 'neon' novae) has been adopted. The impact of this crude approximation on the
evolution of the outburst is extensively discussed in Starrfield et al. (1998).
Results from the two evolutionary sequences are summarized in Table 1. For the purpose of
comparison, we included also results from a third hydrodynamic computation of a  similar model 
(i.e., Model A), for which mixing with material from the 'bare' ONe core is assumed. 
This material is characterized by large amounts of $^{16}$O and $^{20}$Ne, however the specific prescription
depends crytically on details of the previous stellar evolutionary phases. For instance, all the hydrodynamic
models of nova outbursts computed by Starrfield et al. (1998), and references therein,
adopt an ONeMg composition based on C-burning nucleosynthesis by Arnett \& Truran (1969),
which is characterized by large amounts of not only $^{16}$O and $^{20}$Ne, but also $^{24}$Mg, 
with a ratio X($^{16}$O):X($^{20}$Ne):X($^{24}$Mg) = 1.5:2.5:1. In this Letter, as well as in previous
work, we use the most recent prescription for ONe white dwarf composition, for which 
X($^{16}$O):X($^{20}$Ne):X($^{24}$Mg) = 10:6:1 (Ritossa, Garc\'\i a--Berro, \& Iben 1997). 
It is worth noticing that both prescriptions have quite remarkable differences in the amount of
$^{24}$Mg and hence, a significantly different chemical pattern is expected for the ejecta
(in particular, the contribution of novae to the Galactic $^{26}$Al content, a hot topic in
gamma-ray astronomy).

A critical issue in the evolution of the models listed in Table 1 is the initial amount of $^{12}$C
in the envelope: since the main nuclear reaction that triggers the TNR is $^{12}$C(p,$\gamma$), 
models with large initial $^{12}$C abundances (i.e., Model C) will release large amounts of
 energy by CNO-reactions, and hence, will easily achieve the critical conditions to power a TNR.
As a result, the
initial $^{12}$C content determines not only the amount of mass accreted prior to the TNR (and hence, 
the pressure at the base of the accreted envelope), but also the strength of the explosion
(i.e., $\rm T_{peak}$), the amount of mass ejection by the outburst ($\rm \Delta M_{ejec}$), and its 
mean kinetic energy ($\rm K_{ejec}$), as well as the characteristic timescales
of the explosion (accretion phase, rise time to $\rm T_{peak}$, ...). Such differences are reported
in Table 1. Because of the lower amount of energy released by $^{12}$C(p,$\gamma$) reactions
as the initial amount of $^{12}$C decreases, model A accumulates more mass
than models
B and C: the difference between the two extreme cases, model A ---that assumes
mixing with a 'bare' ONe core--- and model C ---that involves material from the thick CO-buffer---
is nearly a factor of two in the accreted envelope mass. 
 Furthermore, both $\rm t_{rise}$ (i.e., time required for the innermost envelope shell to
evolve from a temperature of $3 \times 10^7$ K up to $10^8$ K) and $\rm t_{max}$ (i.e, time
spent by the innermost shell between $10^8$ K and $\rm T_{peak}$) clearly increase as we
decrease the initial amount of $^{12}$C in the envelope. For instance, model  A
takes $\rm t_{rise} = 1.21 \times 10^7$ sec to reach a temperature of $10^8$ K at the base of the envelope 
(also, $\rm t_{max} = 313$ sec), whereas
model C, reaches $10^8$ K much faster, $\rm t_{rise} = 2.5 \times 10^5$ sec (and $\rm t_{max} = 45$ sec). 
Model B can be regarded as an intermediate model between both. 

A direct consequence of the abovementioned differences is that the outburst, whose strength is directly 
determined by the pressure at the base of the envelope (or conversely, by the amount of mass accreted), 
will have a clear imprint in the ejected envelope shells.  A first one, involves
the amount of material ejected by the explosion and its expansion velocity: 
as an example, model A ejects $1.8 \times 10^{-5} \msun$, with a mean
kinetic energy of $\rm K_{ejec} = 1.5 \times 10^{45}$ ergs s$^{-1}$, whereas
 model C ejects $10^{-5} \msun$ (almost a factor of 2 lower), with a mean
kinetic energy of $\rm K_{ejec} = 10^{45}$ ergs s$^{-1}$. A second imprint is found in the accompanying
nucleosynthesis (see Table 1), which is very sensitive to the characteristic timescales of the explosion
as well as to the maximum temperatures reached during the TNR (model A reaches
$\rm T_{peak} = 2.5 \times 10^8 K$, whereas the less violent outburst in  model C reaches only
$\rm T_{peak} = 2.2 \times 10^8 K$). This is, for instance, the case of $^7$Li, strongly enhanced
in the ejecta of model C because of the rapid rise time to temperatures of the order
of $\sim 10^8$ K, where efficient $^8$B($\gamma$,p)$^7$Be photodisintegration reactions prevent
significant destruction of $^7$Be, which is transformed into $^7$Li by electron captures in
the cold, ejected envelope shells (Hernanz et al. 1996).
The larger initial $^{12}$C abundance of this model is also responsible for the moderate increase 
in the final $^{12,13}$C and $^{14,15}$N yields, which translates into 
 a slightly larger (i.e., a factor of 2) $^{14}$N/$^{15}$N isotopic ratio for model C.
A clear imprint of the decrease in the peak temperatures as the initial amount of $^{12}$C
increases can be seen in the increasing final $^{16}$O yield. On the contrary, no significant effect is found
in the final $^{18}$F content and therefore, the expected prompt gamma-ray output at 511 keV and below 
seems to be unaffected by the presence of a thick CO-buffer. 
Furthermore, model C shows a dramatic decline in the final $^{20}$Ne and $^{22}$Na yields, a direct
consequence of the low initial $^{20}$Ne content for this model. Hence, model C would provide
observational features (in particular the lack of significant overabundances of Ne in the spectra) 
typical of a CO novae rather than an ONe one.  However, it is worth mentioning that $^{26}$Al is
not very much affected by the presence of the CO-buffer because of the large amounts of the 'seed' 
$^{25}$Mg both in the transition zone as well as in the buffer itself (see Fig. 1). Therefore,
we can end up with a quite 'weird' nova explosion, ressembling a CO nova, but with simultaneous
ejection of $^7$Li and $^{26}$Al, among additional peculiar features mentioned above.

\section{Discussion and conclusions}

In this Letter, we have shown that classical nova outbursts on top of ONe white dwarfs with thick CO-buffers 
may lead to the ejection of significant amounts of both $^7$Li and $^{26}$Al, usually predicted to
be synthesized in different nova types (CO and ONe novae, respectively). In turn, due to the
lack of significant amounts of neon in the ejecta, such explosions may be (wrongly?) classified
as CO novae.

It is very difficult (and out of the scope of this paper) to estimate the number of 'weird' nova
explosions as compared with the number of normal novae, in particular because the thickness of both
the CO-buffer and the C-rich trasition zone depend critically on the mass of the progenitor star
(see Gil-Pons et al. 2003, for details). However, we have estimated the number of outbursts required
to completely get rid of the CO-buffer and the transition zone, assuming that in each explosion,
50\% of the ejecta corresponds to material eroded from the interior (either from the buffer or from
the transition zone). For this purpose, we adopt the values given in 
Garc\'\i a--Berro et al. (1997),  corresponding to the evolution of a single 9 $\msun$ star, whose
chemical profiles have been adopted in this paper: a thickness of 0.0264 $\msun$ for the CO-buffer
and 0.0236 $\msun$ for the transition zone. Around 7300 outbursts will be required before a nova
with standard composition will show up. 
It has to be emphasized, however, that the discovery of 'neon' novae is a clear evidence that,
at least in some cases, such CO-buffers are lost before the onset of accretion. Otherwise, it would
not be possible to find overabundances of  Ne in the ejecta of such systems and therefore, the 
'neon' (or ONe) nova class would not exist at all.
Therefore, it remains to be analyzed in detail whether such CO-buffers are systematically lost during
the previous evolution or they may remain and give rise to some 'weird' nova explosions.

\acknowledgments

We thank an anonymous referee for valuable comments to the manuscript.
This work has been partially supported by the MCYT grants 
AYA2001-2360 and AYA2002-0494C03-01, by the CIRIT and by the E.U. FEDER funds.

\clearpage
  \begin{figure}
  \plottwo{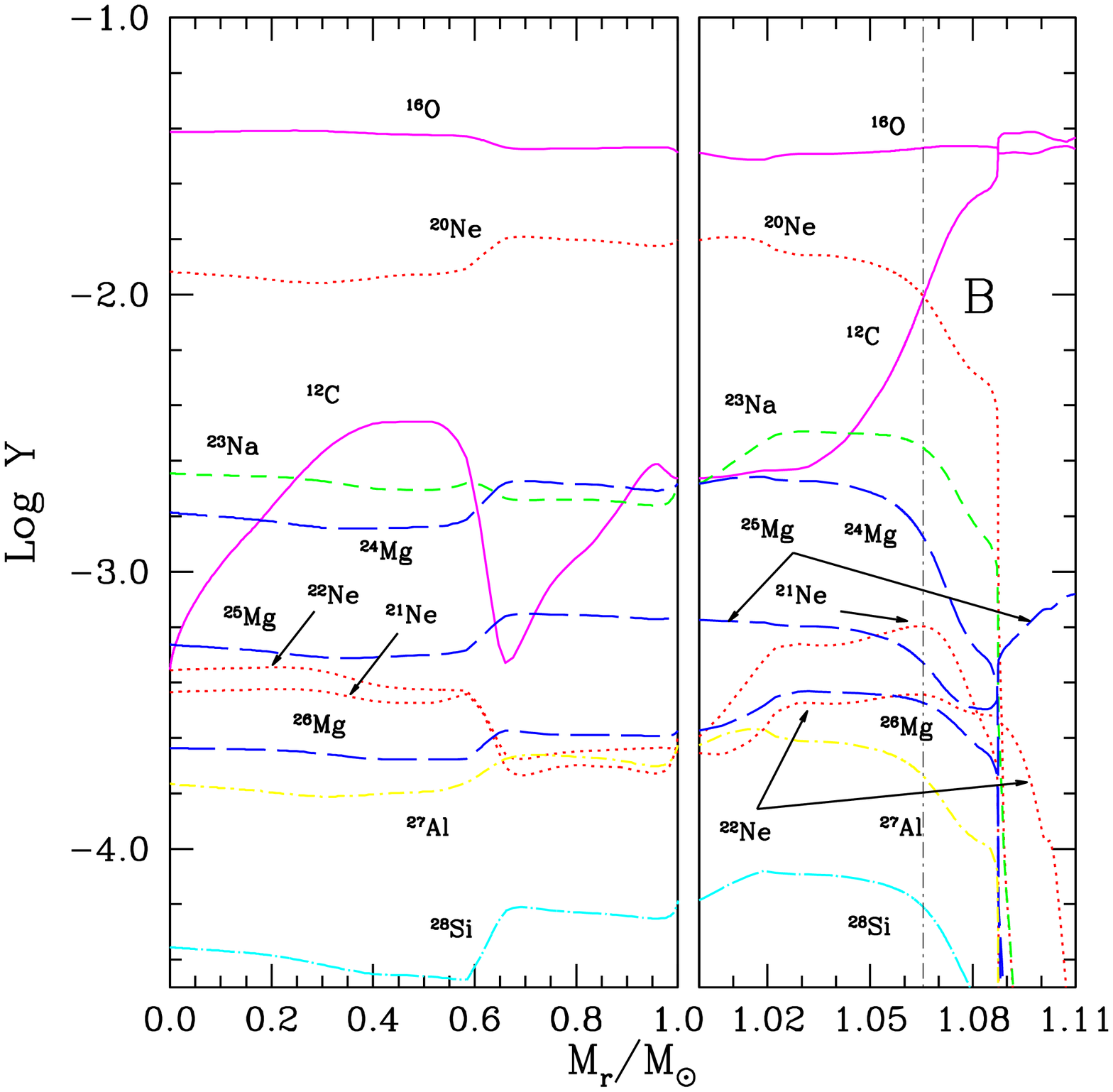}{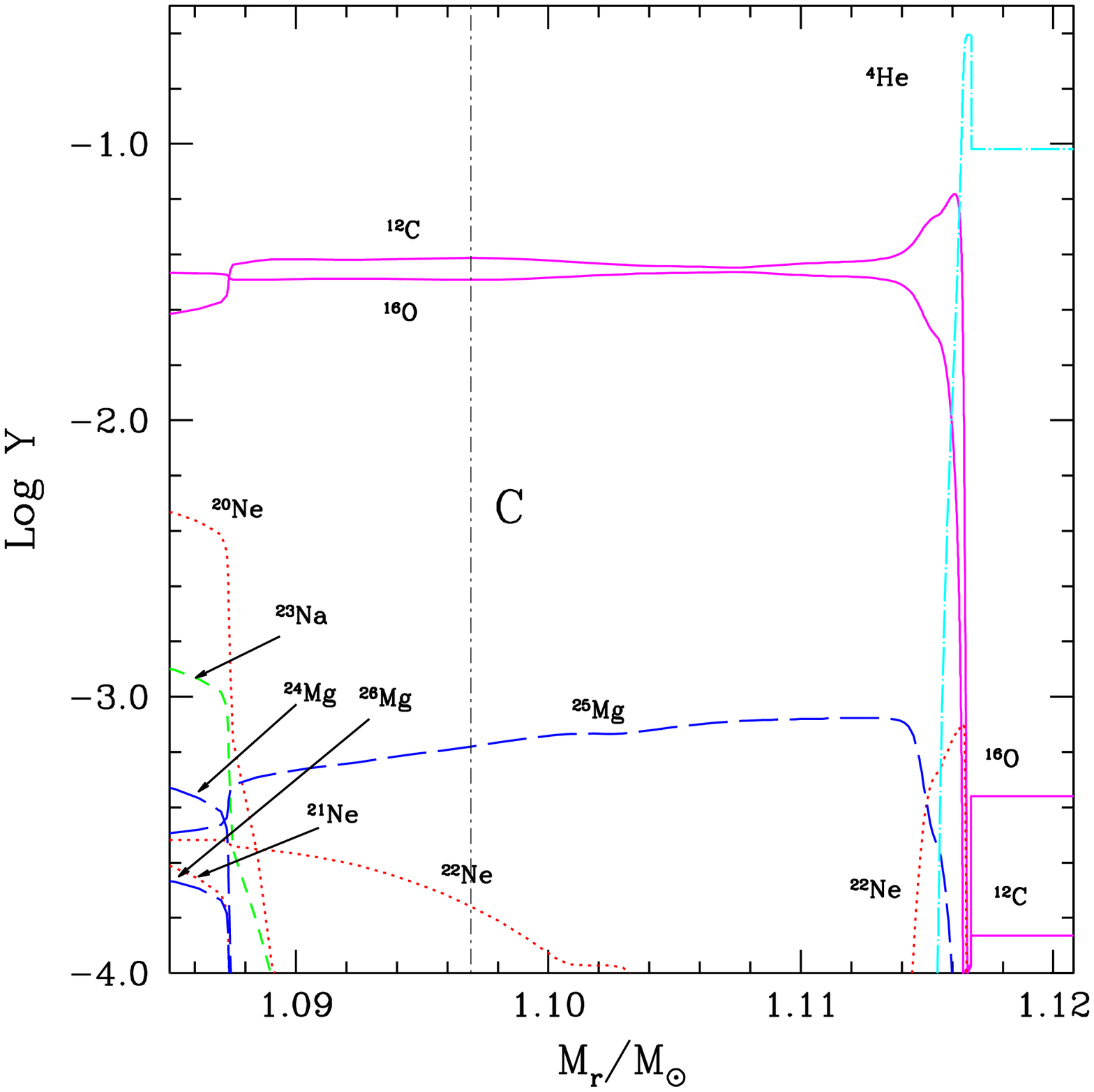}
  \caption{Chemical composition of an ONe white dwarf interior, showing the ONe-rich core
         (Left panel), the CO-buffer on top (Right panel), and the transition zone between
    core and buffer (Middle panel).
   See Garc\'\i a--Berro et al. (1997) for details. The vertical dash dotted lines
    indicate the specific set of abundances adopted in this work.}
  \end{figure}

\end{document}